\documentclass[final,5p,times,twocolumn]{elsarticle}
\usepackage{graphicx}
\usepackage{dcolumn}
\usepackage{amssymb,amsmath,amsthm,mathrsfs}
\usepackage{epstopdf}
\usepackage{lipsum}
\usepackage{hyperref}
\usepackage{empheq}
\usepackage{color}
\usepackage[normalem]{ulem} 
\usepackage{physics}
\usepackage{cleveref}
\usepackage{comment}
\usepackage[ruled,vlined]{algorithm2e}

\newcommand{\be}{\begin{equation}}
\newcommand{\ee}{\end{equation}}
 \newcommand{\bea}{\begin{eqnarray}}
\newcommand{\eea}{\end{eqnarray}}

\definecolor{burntorange}{rgb}{0.8, 0.33, 0.0}

\definecolor{blue_col}{RGB}{0,92,175}
\definecolor{red_col}{RGB}{203,64,66}
\hypersetup{linktocpage,
    colorlinks=true,
    linkcolor=red_col,
    filecolor=Mahogany,      
    urlcolor=blue_col,
    citecolor=blue_col,
}

\begin{document}
\begin{frontmatter}

\title{Adiabatic or Non-Adiabatic? Unraveling the Nature of Initial Conditions in the Cosmological Gravitational Wave Background}

\author{Lorenzo Valbusa Dall'Armi}
\address{Dipartimento di Fisica e Astronomia ``G. Galilei",
Universit\`a degli Studi di Padova, via Marzolo 8, I-35131 Padova, Italy}
\address{INFN, Sezione di Padova,
via Marzolo 8, I-35131 Padova, Italy}

\author{Alina Mierna}
\address{Dipartimento di Fisica e Astronomia ``G. Galilei",
Universit\`a degli Studi di Padova, via Marzolo 8, I-35131 Padova, Italy}
\address{INFN, Sezione di Padova,
via Marzolo 8, I-35131 Padova, Italy}

\author{Sabino Matarrese}
\address{Dipartimento di Fisica e Astronomia ``G. Galilei",
Universit\`a degli Studi di Padova, via Marzolo 8, I-35131 Padova, Italy}
\address{INFN, Sezione di Padova,
via Marzolo 8, I-35131 Padova, Italy}
\address{INAF - Osservatorio Astronomico di Padova, Vicolo dell'Osservatorio 5, I-35122 Padova, Italy}
\address{Gran Sasso Science Institute, Viale F. Crispi 7, I-67100 L'Aquila, Italy}

\author{Angelo Ricciardone}
\address{Dipartimento di Fisica ``E. Fermi'', Universit\`a  di Pisa, I-56127 Pisa, Italy}
\address{INFN, Sezione di Pisa, I-56127 Pisa, Italy}
\address{Dipartimento di Fisica e Astronomia ``G. Galilei",
Universit\`a degli Studi di Padova, via Marzolo 8, I-35131 Padova, Italy}

\date{\today}

\begin{abstract}

The non-thermal nature of the cosmological gravitational wave background (CGWB) makes it difficult to define the initial condition for the graviton distribution, which determines the initial contribution to the CGWB anisotropies. Specifically, unlike Cosmic Microwave Background (CMB) photons, primordial gravitational waves (GWs) are not necessarily subject to adiabatic initial conditions. For the CGWB generated by quantum fluctuations of the metric during inflation the initial conditions are not adiabatic. The violation of adiabaticity arises from the presence of independent tensor perturbations during inflation, which behave as
two extra fields that affect the standard single-clock argument.  The non-adiabatic initial contribution enhances the total CGWB angular power spectrum compared to the standard adiabatic case. Given the high degree of correlation between the CGWB and Cosmic Microwave Background (CMB) anisotropies, we present the constrained realization maps of the CGWB built using the high-resolution CMB Planck maps for different initial conditions. 

\end{abstract}
\end{frontmatter}

\section{Introduction} 
The detection of a stochastic gravitational wave background (SGWB) by the PTA collaboration (i.e., NANOGrav, EPTA/InPTA, PPTA, and CPTA~\cite{NANOGrav:2023gor, Antoniadis:2023ott, Reardon:2023gzh, Xu:2023wog}) opens an exciting era for the study of cosmological (CGWB) and astrophysical (AGWB) backgrounds. Several cosmological or astrophysical interpretations of the detected signal have been proposed in~\cite{NANOGrav:2023hvm,NANOGrav:2023hfp, Antoniadis:2023xlr, Franciolini:2023pbf,Ellis:2023tsl,Franciolini:2023wjm,Vagnozzi:2023lwo,Figueroa:2023zhu,Liu:2023ymk}. Since both the AGWB and the CGWB would retain interesting physical information on the GW sourcing mechanism, it is crucial to have a technique to disentangle the cosmological and the astrophysical contributions with high accuracy. Given the improved angular resolution of future GW detectors, anisotropies in the GW energy density present a new method for differentiating between sources of GWs in the early and late Universe~\cite{Alba:2015cms,Contaldi:2016koz,Cusin:2018rsq, Jenkins:2018nty,Bertacca:2019fnt,Bartolo:2019oiq, Bartolo:2019yeu,ValbusaDallArmi:2020ifo,Mentasti:2020yyd,Bellomo:2021mer,LISACosmologyWorkingGroup:2022kbp,ValbusaDallArmi:2022htu,Mentasti:2023gmg,ValbusaDallArmi:2023ydl}.

Similarly to CMB anisotropies~\cite{Dodelson:2003ft}, it is possible to use a Boltzmann approach to characterize the angular power spectrum of the CGWB~\cite{Contaldi:2016koz, Bartolo:2019oiq, Bartolo:2019yeu, ValbusaDallArmi:2020ifo}, finding that the anisotropies are produced both at production and during the propagation of the waves through the large-scale scalar and tensor perturbations of the Universe. The inhomogeneities in the CGWB density at its production are closely related to the properties of the source, while the redshifting of gravitons during their free streaming is almost entirely constrained by the knowledge of the $\Lambda$CDM parameters, up to some ingredients that influence the evolution of metric perturbations before Big Bang nucleosynthesis~\cite{ValbusaDallArmi:2020ifo,Malhotra:2022ply}. When the CGWB is a direct outcome of the inflaton, which is the case for instance of the decay of the inflaton into GWs during reheating, as it is expected to happen for photons and baryons, the initial conditions are adiabatic in single-field models of inflation, because the features of the energy spectrum of the source are inherited by the decay products. This is consistent with the ‘‘separate universe assumption'', which ensures that when a single-clock sets the evolution of the Universe, the presence of adiabatic modes is unavoidable. However, when the stochastic GWs are generated by the intrinsic quantum fluctuations of the metric, which represent two additional independent degrees of freedom (i.e., the two polarizations of the tensor perturbations), non-adiabatic modes could appear even in single-field inflation. 
Since the quantum fluctuations of the metric and of the inflaton are independent, a rigorous way to compute the initial overdensity of the CGWB, or, equivalently, the intrinsic entropy perturbation of the gravitons is needed. 
Since the energy density of the CGWB is subdominant compared to ordinary matter, gravitational radiation plays a negligible role in Einstein's equation and hence the only way to compute the initial conditions is to perturb the energy-momentum tensor defined in terms of the gravitational strain. These ‘‘Inflationary Initial Conditions'' (IIC) lead to an enhancement of the CGWB angular spectrum by about one order of magnitude. The detailed computation is provided in \cite{ValbusaDallArmi:2024hwm}.

The CGWB is expected to correlate with other cosmological observables such as the CMB, since their anisotropies originate from the same source, that is the large-scale scalar and tensor perturbations of the metric.  We analyze the effect of the non-adiabatic initial conditions on the cross-correlation between CGWB and CMB anisotropies and build constrained realization maps of the CGWB extracted from the high-resolution CMB Planck maps in the low noise regime \cite{Ricciardone:2021kel}.

\section{Boltzmann formalism for gravitons}
We consider a perturbed Friedmann-Lemaître-Robertson-Walker (FLRW) spacetime, whose line element in the Poisson gauge reads
\begin{equation}
ds^2 = a^2(\eta)\left[-e^{2\Psi}d\eta^2 +e^{-2\Phi}\left(e^{\gamma}\right)_{ij}dx^idx^j\right],
\end{equation}
where $\Phi$ and $\Psi$ are the large-scale scalar perturbations and $\gamma_{ij}$ is the transverse-traceless tensor perturbation. The latter can be decomposed into the small-scale $h_{ij}$ and the large-scale modes $H_{ij}$, as $\gamma_{ij}\equiv h_{ij}+H_{ij}\,$. Our observable for interferometers is the CGWB generated by the tensor perturbation $h_{ij}$, whose comoving momentum $q$ is taken to be much larger than the typical comoving scales $k$ over which the background varies. 
Therefore, it is possible to use the shortwave approximation~\cite{Isaacson:1967sln}, which allows to describe the propagation of GWs by using geometric optics in a curved manifold, where the geometry is purely determined by the background quantities and by the large-scale perturbations (i.e., $\Phi$, $\Psi$ and $H_{ij}$). Therefore, it is possible to introduce a distribution function $f_{\rm GW}(\eta,\Vec{x},\Vec{q})$ for gravitons and to compute its evolution by using the collisionless Boltzmann equation~\cite{Alba:2015cms,Contaldi:2016koz,Bartolo:2019oiq,Bartolo:2019yeu}, because the collision term is a higher-order term in the Planck mass~\cite{Misner:1973prb,Bartolo:2018igk}. The homogeneous and isotropic contribution to the distribution function $\bar{f}_{\rm GW}(q)$ solves the Boltzmann equation at zeroth order and it is sensitive just to the expansion of the Universe, while the perturbation of the distribution function, which can be recast by using $\delta f_{\rm GW} = -q\, d\bar{f}_{\rm GW}/dq \, \Gamma$, where $\Gamma$ is the GW analog of the fractional perturbation of the CMB temperature, is the sum of an initial, of a scalar and of a tensor contribution,
\begin{equation}
    \Gamma(\eta_0,\Vec{k},\Vec{q}) = \Gamma_I(\eta_0,\Vec{k},\Vec{q})+\Gamma_S(\eta_0,\Vec{k})+\Gamma_T(\eta_0,\Vec{k}) \, ,
\end{equation}
with $\eta_0$ the present time. The scalar contribution takes into account the redshifting of gravitons due to the metric perturbations, including the Sachs-Wolfe (SW) and the Integrated Sachs-Wolfe (ISW) effects~\cite{Bartolo:2019oiq, Bartolo:2019yeu}, similarly to the CMB case ~\cite{Sachs:1967er,White:1994sx}, while the tensor part contains an analogous ISW effect for tensors~\cite{Pritchard:2004qp}. The first numerical computation of the angular power spectrum of the CGWB has been done in~\cite{ValbusaDallArmi:2020ifo}, while a public Boltzmann code \texttt{GW\_CLASS} has been recently developed~\cite{Schulze:2023ich}. In this work, we focus on the initial condition contribution to the solution of the Boltzmann equation,
\begin{equation}
    \Gamma_I(\eta_0,\Vec{k},\Vec{q}) = e^{i\Vec{k}\cdot \hat{q}(\eta_0-\eta_{\rm in})}\Gamma(\eta_{\rm in},\Vec{k},\Vec{q}) \, , 
\end{equation}
with $\eta_{\rm in}$ the initial time at which we compute the initial conditions. The connection between the perturbation of the distribution function $\Gamma$ and the perturbation of the energy density in a given frequency bin, which is the observable at interferometers, is given by $\delta_{\rm GW}(\eta,\Vec{x},\Vec{q}) = [4-n_{\rm gwb}(q)]\Gamma(\eta,\Vec{x},\Vec{q})$, with $n_{\rm gwb}$ the tensor tilt of the cosmological background at the frequency $f = q/2\pi$~\cite{Bartolo:2019oiq,Bartolo:2019yeu}. In the CMB case, the initial condition is set at the time when photons are tightly coupled with baryons, therefore any deviation from a thermal spectrum would be suppressed and the perturbation of the distribution function is just the fractional fluctuation of the CMB temperature, $\delta T(\hat{n})/T$, which does not depend on the magnitude of the photon momentum. Since the CGWB is produced at much higher temperatures, it is necessary to discuss the initial conditions more carefully, because the decoupling of gravitational interactions implies a non-trivial spectrum and a perturbation of the distribution function that could depend on frequency.
If the CGWB is produced by the inflaton decay together the other particle species that fill the universe, then the initial conditions for the energy density of the CGWB are connected to the perturbations of the inflaton. In the case of a CGWB produced by quantum fluctuations of the metric during inflation, the situation is different. In this case, in order to use the shortwave approximation, the initial conditions for the CGWB are set long after the horizon crossing of the high-frequency GWs and the amount of energy produced as gravitational radiation is subdominant compared to standard radiation. Since Einstein's equations are almost insensitive to the presence of the CGWB, because its energy density is subdominant, (i.e., $\bar{\rho}_{\rm GW}/\bar{\rho}_{\rm rad}\ll 1$) the $(0,0)$ component carries almost no information on $\delta_{\rm GW}$.  

\section{Inflationary initial conditions for the CGWB}
The definition of the energy-momentum tensor of the GWs is ambiguous when the wavelength of the waves is comparable to the scales over which the background metric varies~\cite{Giovannini:2019ioo} and different definitions can be used~\cite{Ford:1977dj,Ford:1977in,Landau:1975pou,Babak:1999dc}. Initial conditions are set when the GWs are well inside the horizon, because it is possible to use the shortwave approximation and geometric optics, describing gravitons in terms of a distribution function whose evolution is governed by the Boltzmann equation. When the GWs are well inside the causal horizon, all the definitions of the energy-momentum tensor converge~\cite{Isaacson:1967sln,Misner:1973prb,Landau:1975pou,Giovannini:2019ioo}. Therefore, we can use a simple definition of the energy-momentum tensor in terms of covariant derivatives of the radiative degrees of freedom of the metric~\cite{Isaacson:1967sln}
\begin{equation}
    T_{\mu\nu}^{\rm GW} = \frac{1}{32\pi G}\left\langle\mathcal{D}_\mu \gamma_{\alpha\beta}^{\rm GW}\mathcal{D}_\nu\gamma^{\rm GW\,\alpha\beta}\right\rangle \, ,
    \label{def:T_mu_nu_Isaacson}
\end{equation}
where the radiative degrees of freedom of the metric in
the Poisson gauge are non-vanishing only for the spatial indices, $\gamma_{ij}^{\rm GW} = a^2 h_{ij}$ and\footnote{Here round parentheses denote symmetrization over indices.} $\gamma^{{\rm GW}\, ij} = -[(1+4\Psi)h^{ij}+H^{k(i}h^{j)}_k]/a^2$. The result can also be obtained by using the prescription of~\cite{Isaacson:1967sln} with the approach of~\cite{Landau:1975pou}, in which the energy-momentum tensor is obtained by perturbing the Einstein tensor up to ‘‘hybrid third order'' in the perturbations. In both cases, we keep the terms quadratic in $h_{ij}$ and up to linear order in the large-scale scalar $\Phi, \Psi$ and tensor $H_{ij}$ perturbations. The energy density perturbation can be related to the perturbation of the graviton distribution function
\begin{equation}
\begin{split}
       \Gamma(\eta_{\rm in}, k)  = &\frac{1}{4-n_{\rm gwb}(q)} \left[4\Psi(\eta_{\rm in},\Vec{k})-2\Phi(\eta_{\rm in},\Vec{k}) \right.\\&\left.+2 H(\eta_{\rm in},\Vec{k})\left(1-\mu^2\right)\right]\, ,  
     \label{eq:Gamma_I_IIC}
\end{split} 
\end{equation}
where $\mu = \hat{k}\cdot \hat{q}$ and we have considered an unpolarized GW background $H_{ij}(\eta,\vec{k}) = \sum_\lambda H(\eta,\vec{k})e^\lambda_{ij}(\hat{k})$. These inflationary initial conditions (IIC) contain an additional contribution compared to the standard adiabatic initial conditions (AD)
\begin{equation}
\begin{split}
     \Gamma(\eta_{\rm in}, k) &=- \frac{2}{4-n_{\rm gwb}(q)}\Psi(\eta_{\rm in},\Vec{k})\, ,  
     \label{eq:Gamma_I_IIC}
\end{split} 
\end{equation}
Expanding the field in spherical harmonics, the initial contribution to the graviton distribution function can be written as
  \begin{equation}
  \begin{split}
  \Gamma_{\ell m,I}  =& 4\pi (-i)^{\ell} \\
  &\int \frac{d^3k}{(2\pi)^3}  \Bigl[  Y^*_{\ell m}(\hat{k})\mathcal{R}(\Vec{k})\Delta^{I-S}_\ell(\eta_0,k,\eta_{\rm in}) \\
   &\hspace{4em}+\sum_\lambda Y^{*\, -\lambda}_{\ell m}(\hat{k}) H_\lambda(\Vec{k})\Delta^{I-T}_\ell(\eta_0,k,\eta_{\rm in})  \Bigl] \, ,
  \end{split}
  \end{equation}
where $\mathcal{R}$ is the gauge-invariant primordial curvature perturbation, $\Delta^{I-S}_\ell$ and $ \Delta^{I-T}_\ell$ are the source functions given by 
\begin{equation}
    \begin{split}
    \Delta^{I-S}_\ell = & \frac{4T_\Psi(\eta_{\rm in},k)-2T_\Phi(\eta_{\rm in},k)}{4-n_{\rm gwb}(q)}j_\ell[k(\eta_0-\eta_{\rm in})] \, , \\
    \Delta^{I-T}_\ell = & -\frac{T_H(\eta_{\rm in},k)}{4-n_{\rm gwb}(q)}\frac{1}{2}\sqrt{\frac{(\ell+2)!}{(\ell-2)!}}\frac{j_\ell[k(\eta_0-\eta_{\rm in})]}{k^2(\eta_0-\eta_{\rm in})^2} \, .
    \end{split}
    \label{eq:source_function_IC}
\end{equation}
The transfer functions $T_\Phi$, $T_\Psi$, $T_H$ describe the time evolution of the metric perturbations, while the primordial information is contained in $\mathcal{R}$ and $H_\lambda$ (see~\cite{Bartolo:2019oiq,Bartolo:2019yeu,Schulze:2023ich} for more details). We consider a statistically isotropic signal. The angular power spectrum of the CGWB then is defined as 
\begin{equation}
       \langle \delta_{{\rm GW},\ell m}\delta^*_{{\rm GW},\ell'm'} \rangle = \delta_{\ell \ell'}\delta_{mm'}C_{\ell }^{\rm CGWB}  \,,
\end{equation}
and it is the sum of a contribution given by the initial conditions and a contribution given by the propagation through the scalar and tensor perturbations,  
  \begin{equation}
  \begin{split}
  \frac{C^{\rm CGWB}_{\ell}}{4\pi(4-n_{\rm gwb})^2} =&  \int \frac{dk}{k} \Bigl[
   P_{\mathcal{R}}(k) \left(\Delta^{I-S}_\ell+\Delta_\ell^{\rm SW}+\Delta_\ell^{\rm ISW}\right)^2 \\
   &+ P_{T}(k) \left(\Delta^{I-T}_\ell+\Delta_\ell^{{\rm ISW}-T}\right)^2\Bigl] \, ,
   \end{split}
  \end{equation}
where $P_{\mathcal{R}}(k)$ and $P_{T}(k)$ are the primordial scalar and tensor spectra respectively. 
\section{Constrained realizations maps of CGWB}
In the geometric optics limit, gravitons and CMB photons propagate along the same perturbed geodesics. For this reason, the angular power spectrum of the CGWB and the CMB are highly correlated at large angular scales, where both are dominated by the initial condition contribution and the SW effect. Based on Planck constraints, the initial conditions for CMB photons are adiabatic\cite{Planck:2018vyg}. The solution of Boltzmann equation for the photon graviton distribution functions expanded in spherical harmonics is given by  the sum of the adiabatic initial conditions, the SW, Doppler and ISW contributions  
  \begin{equation}
  \begin{split}
  \theta_{\ell m}  &= 4\pi (-i)^{\ell} \int \frac{d^3k}{(2\pi)^3} Y^*_{\ell m}(\hat{k})\\&\int^{\eta_0}_{\eta_{i }} \, d\eta\left[g(\eta)(\Theta_0 (\eta,k)+ \Phi(\eta,k))j_\ell[k(\eta_0-\eta)] \right.\, \\
   &\left.+ g(\eta) \frac{-iv_b(\eta,k)}{k}\frac{d}{d\eta} j_\ell[k(\eta_0-\eta)] \right.\\&\left. + e^{-\tau(\eta)}\left(\Phi^\prime (\eta,k) + \Psi^\prime (\eta,k)\right) j_\ell[k(\eta_0-\eta)] \right]\,,
  \end{split}
  \end{equation}
where  $\tau$ is the optical depth, $\tau(\eta)= \int^{\eta_0}_{\eta}d\eta^\prime \, n_e \sigma_T a$ with $n_e$ the free
electron number density and $\sigma_T$ the Compton cross-section, $v_b$ denotes the velocity of baryons, and $g(\eta)$ the visibility function \cite{Dodelson:2003ft}. We have neglected the tensor contribution to the CMB anisotropies through the ISW effect, since it is subdominant w.r.t. the scalar one. We will also disregard the tensor contribution to the CGWB anisotropy when generating the constrained realization maps. 

The angular power spectrum of the CMB and the cross-correlation with the CGWB can be defined correspondingly 
\begin{equation}
    \begin{split}
         \langle \theta_{\ell m}\theta^*_{\ell'm'} \rangle =& \delta_{\ell \ell'}\delta_{mm'}C_{\ell }^{\rm CMB}  \, ,\\
          \langle \delta_{{\rm GW},\ell m}\theta^*_{\ell'm'} \rangle =& \delta_{\ell \ell'}\delta_{mm'}C_{\ell }^{\rm CGWB\times CMB}\, ,
    \end{split}
\end{equation}
\begin{figure*}[t]
    \centering
    \includegraphics[width=.32\linewidth]{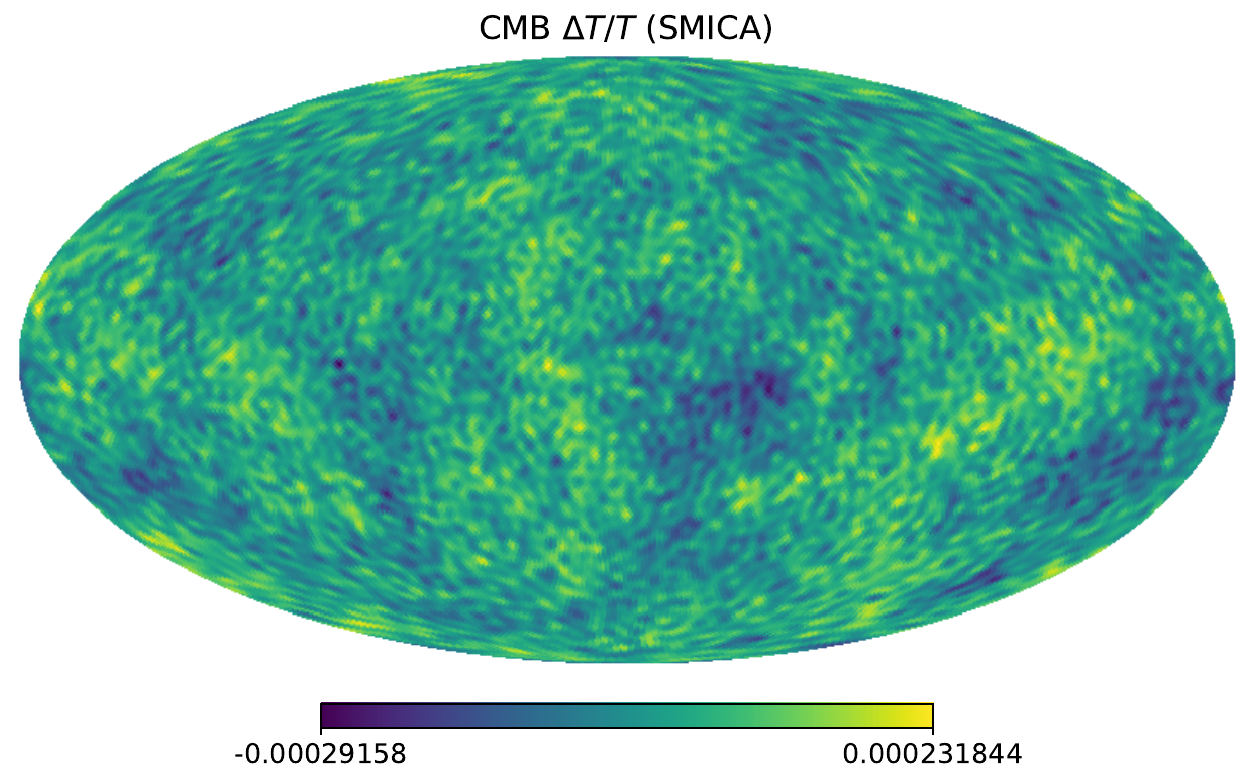}
    \includegraphics[width=.32\linewidth]{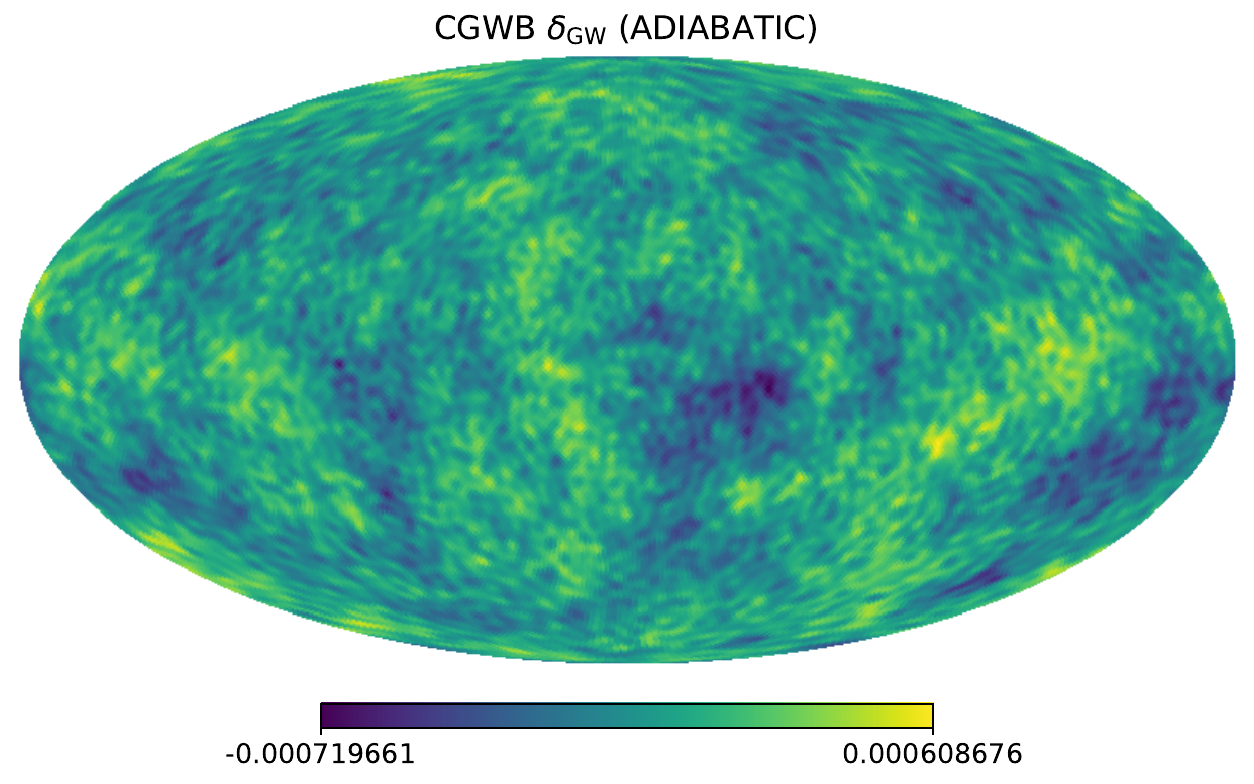}
    \includegraphics[width=.32\linewidth]{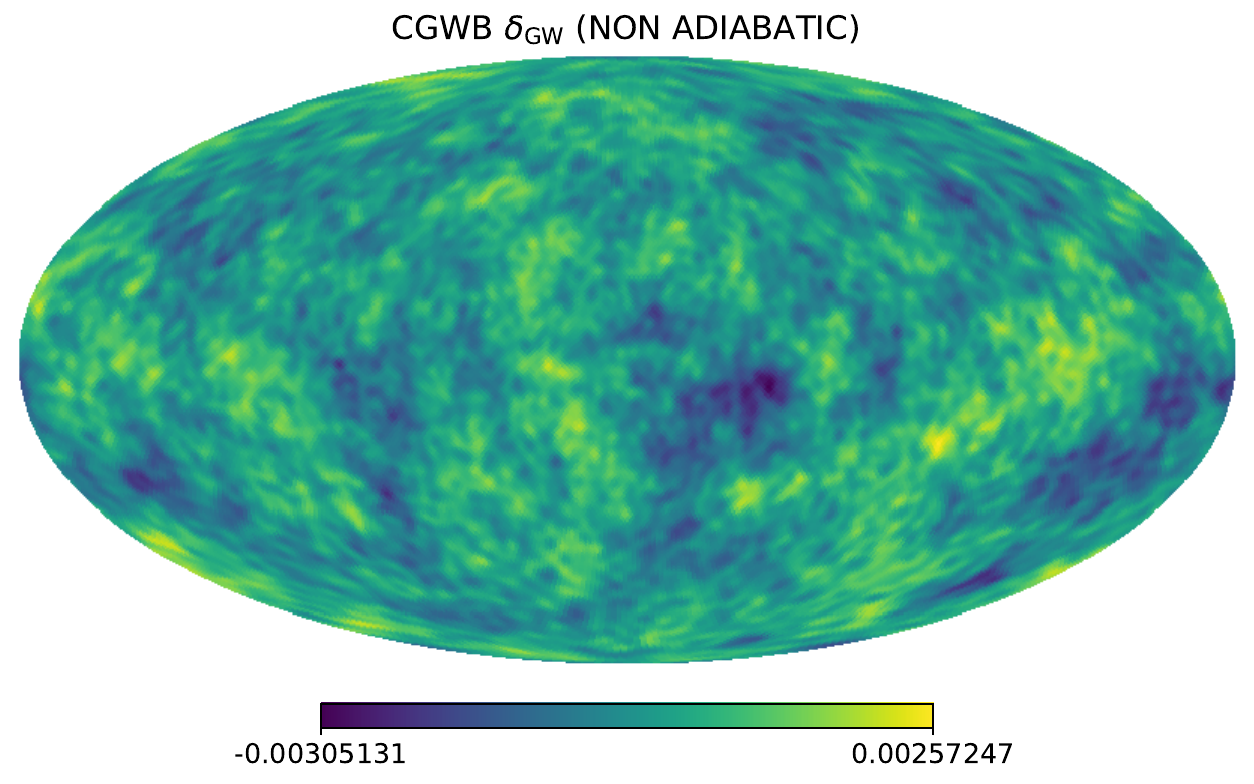}
 
    \caption{Left plot: Planck CMB SMICA map with $\ell_{\rm max} =200$ Central plot: constrained CGWB map with adiabatic initial conditions (AD) with $\ell_{\rm max} =200$ in the noiseless case; Right plot: constrained CGWB map with inflationary initial conditions (IIC) with $\ell_{\rm max} =200$ in the noiseless case.} 
    \label{fig}
\end{figure*}
and correlation coeﬃcient $r$ between the CMB and the
CGWB is  
\begin{equation}
    r_\ell^{\rm CMB\times GW } \equiv \frac{C_\ell^{\rm CMB\times GW}}{\sqrt{C_\ell^{\rm CMB}C_\ell^{\rm GW}}} \, .
\end{equation}
One can show that, on large scales, for AD we get $r\approx 0.98$, while in the case of IIC we find $r\approx 0.8-0.9$.
Therefore, measuring such a correlation would provide a way to test the nature of the initial conditions once a CGWB is detected.

Since the correlation between the CMB and the CGWB is close to 1 on large scales, we can use the CMB observations to generate constrained realization maps of the CGWB \cite{Ricciardone:2021kel}. The spherical harmonics coeﬃcients $a_{\ell m}$ and $\Gamma_{\ell m}$ are distributed as Gaussian random variables with zero average and covariance given by a block-diagonal matrix, with 2 blocks, with each block equal to 
\begin{equation}
C_{\ell}^{\rm block} =
    \begin{pmatrix}
C_{\ell}^{CGWB} & C_{\ell}^{\rm CMB \times CGWB} \\
C_{\ell}^{\rm CMB \times CGWB} & C_{\ell}^{\rm CMB} 
\end{pmatrix}\,,
\end{equation}
The conditional probability of $\Gamma_{\ell m}$ given that $a_{\ell m}$ is a Gaussian with mean and elements of the covariance matrix given by
\begin{equation}
\begin{split}
     \mu_{\ell m} =& \frac{C_{\ell}^{\rm CMB \times CGWB}}{C_{\ell}^{\rm CMB}}a_{\ell m} \,, \\  \Sigma_{\ell m} = &C_{\ell}^{\rm CGWB} - \frac{(C_{\ell}^{\rm CMB \times CGWB})^2}{C_{\ell}^{\rm CMB}}\, .
\end{split}
\end{equation}
The coefficients $a_{\ell m}$ can be extracted from the full-sky, Planck SMICA CMB map. When r approaches 1, we can predict the hot and the cold spots of the CGWB map from the CMB map, since due to the large degree of correlation the covariance of the conditional Gaussian goes to zero.  On small scales, the map would be modified by different realizations of the system due to the uncertainty.

In Figure \ref{fig} we present a realisation of the constrained CGWB map with mean $\mu_{\ell m}$ and covariance $\Sigma_{\ell m} $ for adiabatic and inflationary initial conditions considering the Planck best-fit parameters \cite{Planck:2018vyg}. In the non-adiabatic case, the map appears to have a lower effective angular resolution. This is due to the fact that the contrast between the large- and small-scale angular power spectrum is reduced compared to the adiabatic case. As a result, the map is not dominated by small-scale features, but shows a more comparable contribution from large angular scales.

\section{Conclusions} 
The CGWB initial anisotropy strongly depends on the specific mechanism of GW production. The inflationary initial conditions lead to an amplification of the total CGWB angular power spectrum when compared to the standard adiabatic case. This enhancement could play a crucial role in the detection of the anisotropies of the cosmological background in presence of intrinsic and instrumental noise~\cite{Mentasti:2023icu,Mentasti:2023gmg}. The inclusion of non-adiabatic initial conditions may also increase the sensitivity of the anisotropies to early-universe fundamental observables (e.g., non-Gaussianity~\cite{Perna:2023dgg}). 
The cross-correlation of the CGWB and the CMB on large scales offers another way to distinguish among different sources of GWs in the early Universe and provides a new observable for extracting cosmological information. We have obtained constrained realization maps of the CGWB from the high-resolution Planck CMB map for inflationary initial conditions and adiabatic initial conditions.
\section{Acknowledgments.}
\noindent
We thank N. Bartolo, D. Bertacca, I. Caporali, and A. Greco for useful discussions. S.M. acknowledges partial financial support by ASI Grant No. 2016-24-H.0. A.R. acknowledges financial support from the Supporting TAlent in ReSearch@University of Padova (STARS@UNIPD) for the project “Constraining Cosmology and Astrophysics with Gravitational Waves, Cosmic Microwave Background and Large-Scale Structure cross-correlations’'.
\vskip 1.5cm

\bibliographystyle{elsarticle-num} 
\bibliography{Biblio}

\end{document}